# COSMO-PHYSICAL EFFECTS IN THE TIME SERIES OF THE GCP NETWORK

S.E. Shnoll[1,2], V.A. Panchelyuga[2]

Moscow State University, Moscow, Russia (1),
Institute of Theoretical and Experimental Biophysics RAS, Pushchino, Russia (2).
*shnoll@iteb.ru, panvic333@yahoo.com*

In the GCP network, an Internet system of noise generators developed under the direction of Prof. R. Nelson and deployed at various geographical points, synchronous measurements of *a priori* random noise processes are carried out every second [1, 2]. The time series obtained in these measurements are "quite random" from the viewpoint of traditional methods of statistical analysis. However, as shown in our works [3-13], using the method of histogram comparison allows one to discover clear regularities in the noise processes of any nature, from biochemical reactions to radioactive decay or noise in the gravitational antenna. These regularities cannot be revealed by the traditional methods of time series analysis and seem to be determined by fluctuations of the space-time, which result from the movement of the "laboratory" relative to the mass thicknesses (celestial bodies) [4-6, 13]. The application of this method to the analysis of time series of the GCP system showed all the main regularities described earlier: the histogram pattern changes with well-resolved sidereal and solar daily periods; at different geographical points, similar histogram patterns are highly probable to appear at the same local time; histograms with a specific "eclipse" pattern appear synchronously "all over the Earth" at the moments of culmination of a solar eclipse.

1. INTRODUCTION.

The initial material for our studies of the effect of macroscopic fluctuations is time series of amplitude fluctuations obtained by measuring parameters of processes of various nature. Their analysis by the method of histogram analysis [3,4] revealed a number of periods relating to the increased probability of histograms of a certain pattern to appear [11,13]. It was also found that the probability of similar histogram patterns to appear increases synchronously by local time [13]. To investigate the effect of synchronization by local time we need to conduct measurements at different geographical points. In this connection, appearing very promising were time series obtained with a world-wide network of physical random number generators, which was deployed within the framework of the GCP project. The first analyses performed in 2000 yielded reassuring results, but the further acquaintance with the working algorithm of GCP generators cast serious doubts upon the applicability of GCP series to the study of our effects. They were due to the use of the "exclusive OR" operation, i.e. XOR-mask, on the output of a physical random number generator. In this connection, the use of GCP series in the studies of the effect of macroscopic fluctuations was suspended.

In the present work, a possibility to use data of the GCP monitoring for the study of the effect of macroscopic fluctuations has been demonstrated. We have revealed the major regularities, characteristic for the phenomenology of the effect of macroscopic fluctuations, and have analysed the distortions caused by the XOR-mask in the pattern of histograms.

## 2. A BRIEF DESCRIPTION OF THE GCP NETWORK.

At present, the GCP project unites more than 60 computer random number generators (RNGs) placed at different points of the globe [1]. Each of them performs continuous monitoring of a noise process, which depends on the type of the physical process used in this RNG. Generators are connected, through Internet, to a server in Princeton. It performs functions of data collection and processing and also maintains a database, being continuously filling since 1988 [1]. All RNGs are synchronized through Internet; the data synchronization occurs periodically (once per second) all over the network. After synchronization, data are transferred to the server in Princeton, where they are recorded in the database with open access [2].

There are three types of RNGs used in the GCP project. In general, they have the same construction and differ by the type of the source of white noise. The following sources of noise are used:
1) thermal noise in a resistor;
2) noise in a field-effect transistor;
3) noise in Zener diodes.

The working algorithm of a typical generator of the GCP network is the following. The analog signal from an RNG, having a uniform spectrum on the frequency band from 1100 Hz to 30 kHz, goes to a low-pass filter with a cut-off frequency of 1000 Hz, which removes frequencies below the digitizing rate. After amplification, the low-frequency signal is transformed into a meander, allowing one to continue work with a digital representation of the signal.

The digital processing of a signal is carried out by applying a XOR-mask, which performs the "exclusive OR" operation over the meander obtained and the impulses of a clock generator with a frequency of 1000 Hz. After the XOR-mask, the number of bits of the output sequence, corresponding to 200 periods of the clock generator, is counted. The resulting value represents the output signal of the RNG. It is recorded into the database in Princeton and stored there as a point in a time series.

## 2. MATERIALS AND METHODS.

In the analysis of time series obtained from GCP generators, we used methods, which were elaborated and used by us during many decades [3-5,7]. The methods are based on the pairwise comparison of similarity of patterns of smoothed histograms constructed for non-overlapped short segments of time series. All procedures of constructing, scaling, smoothing etc. are performed with a computer program *Histogram Manager* by Edwin Pozharsky. However, the decision about similarity of two histograms is made by an expert based on a visual evaluation. We have to do expert evaluation since there is no effective computer algorithms for comparing histogram patterns, which would be adequate to the integral evaluation by a human-expert. These problems were considered in detail earlier [3,4,13].

A remarkable property of histograms is independence of their pattern of the alteration order of results of measurements within the segment of a time series taken to construct a histogram. For example, for each histogram from those usually used in our works (constructed from 60 points of a time series), there are about $10^{82}$ of different realizations of a 60-point segment of a time series, which keep the histogram pattern invariant. The histogram pattern reflects the spectrum of amplitude fluctuations of a measured quantity within a given segment of a time series. As we showed, the shape of this spectrum is determined by cosmo-physical causes [13].

The regularities in the change of histogram pattern do not depend on the nature of a process and can be found in "absolutely random", from the viewpoint of traditional criteria, time series. This was confirmed by our many years investigation of time series obtained by measuring fluctuations of various processes [3-13].

## 3. RESULTS.

Using our methods of analysis, we have found all major phenomena of "macroscopic fluctuations" in the time series of the GCP network. This means that the cosmo-physical laws, which earlier were shown to be inherent in any physical stochastic processes, are revealed in the system of the GCP generators.

The figures presented below are to demonstrate that the main regularities obtained upon studying the time series of various physical processes are identical to the regularities found for the time series generated by the RNGs of the GCP network.

Fig. 4*a* shows a time series, which is a direct result of registration of noise from generator No. 28 (Roger Nelson, Princeton, NJ, USA, 40°35′ latitude north and 74°65′ longitude west) of April 8, 2005. Fig. 4*b* represents a distribution function constructed on the basis of the signal of Fig. 4*a*. One can note the complete absence of any trends and clear correspondence to "white noise".

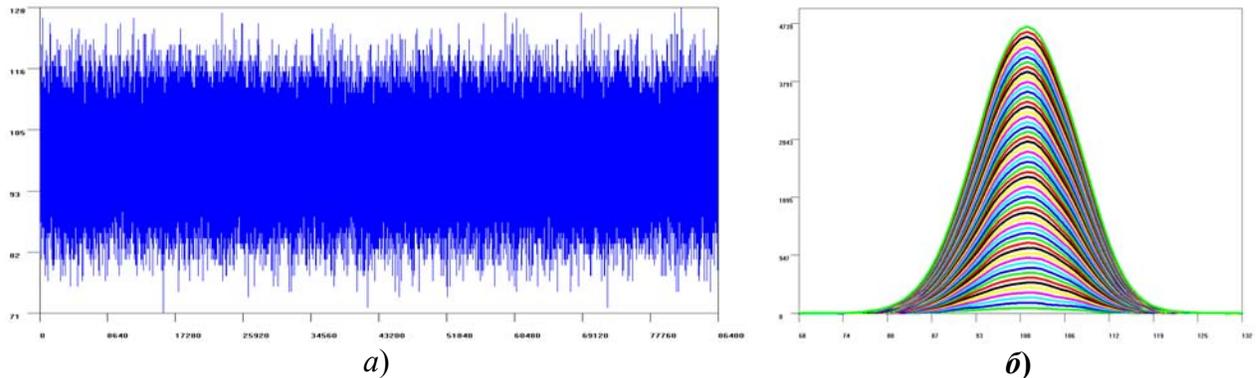

*a)*  *б)*
Fig. 4. A segment of a time series obtained from GCP generator no. 28 of April 8, 2005 (*a*) and its distribution function (*b*).

According to the standard method for our research [3,4,7], the time series (Fig. 4*a*) was divided to non-overlapping segments, with 30 measurements in each of them. For every such segment, a histogram was constructed, which then was smoothed with a 4-point square window for 3 times. Each histogram thus corresponds to a 30-second segment of the source time series. All further work was carried out with these histograms saved in a computer archive of Histogram Manager. Fig. 5 shows a fragment of this journal, which illustrates a sequence of 35 smoothed histograms, corresponding to the successive 30-point segments of the time series.

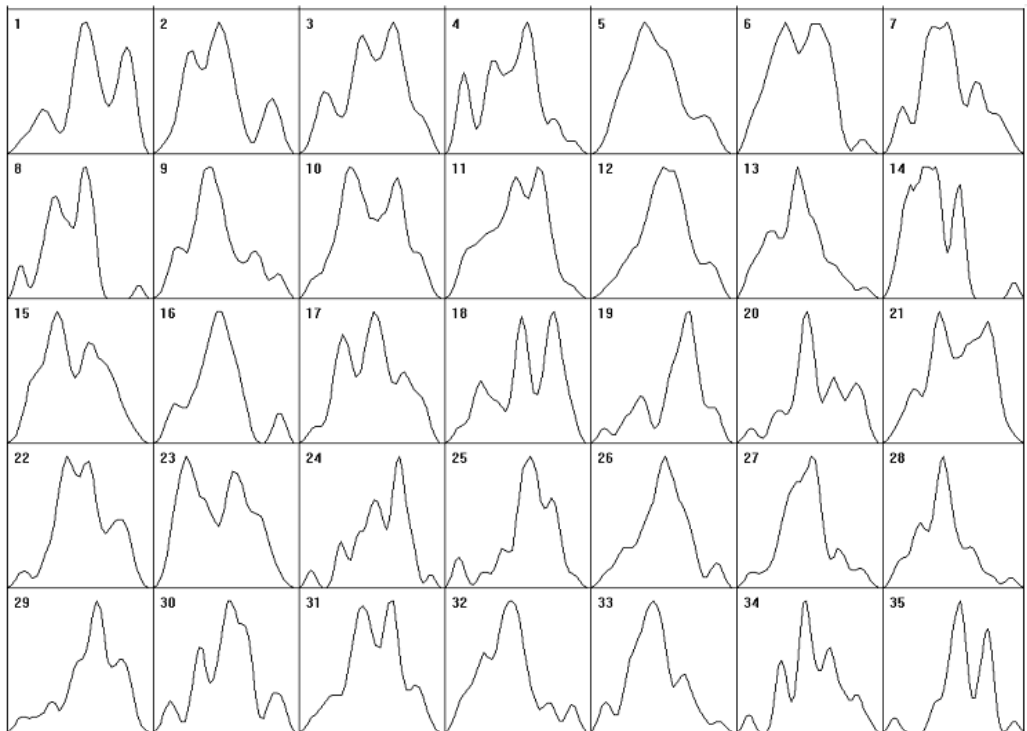

Fig. 5. A fragment of the computer journal. The order numbers of histograms correspond to the successive 30-point segments of the source time series.

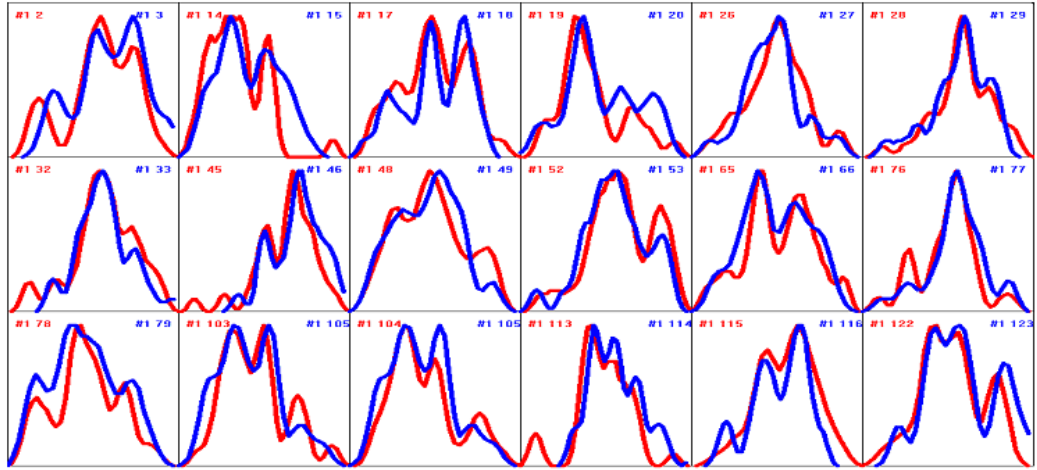
Fig. 6. A fragment of the computer archive: examples of pairs, in which histograms are qualified by an expert as similar.

The main method of further analysis is pairwise comparison of histogram patterns and construction of a distribution of the number of matches depending on the duration of the time interval between them. To obtain significant distributions one needs to perform thousands of pairwise comparisons. In the process of comparison, the expert is allowed to do linear transformations of the histogram pattern: stretching/squeezing, shifting, mirror transformation.

Given in Fig. 6 are examples of pairs of neighbour histograms, which are qualified to be similar upon the expert evaluation. This figure also illustrates the above-mentioned difficulty in algorithmization of the process of histogram comparing. An expert would easily evaluate the similarity of the "idea of form", while various algorithmic estimations appear to be of little use [5,13].

**1) The effect of "near zone" is the first piece of evidence for the cosmo-physical conditionality for the histogram patterns in the time series of the GCP system.**

The first and most easily revealed manifestation of "macroscopic fluctuations" is the effect of "near zone". It means a significantly higher probability of near, neighbour histograms to be similar. This effect can be clearly seen in the time series created by GCP generators. An example is given in Fig. 7. It demonstrates how the number of similar pairs of 0.5-min histograms constructed from the data of Fig. 3 changes with the increase of the time interval between them.

Fig. 7*a* shows that in a set of 700 histograms (so was the analysed segment of a data series), 80 pairs of the nearest neighbours (i.e. with the time interval of 0.5 min) are found to be similar, and for each of other, longer intervals, the number of similar pairs does not exceed 32. In Fig. 7*b*, the same distribution is shown with the values of mean-square errors. The effect of "near zone"

implies the existence of a general external cause, which determines the histogram pattern. And this suggests that the fine structure of histograms is not casual.

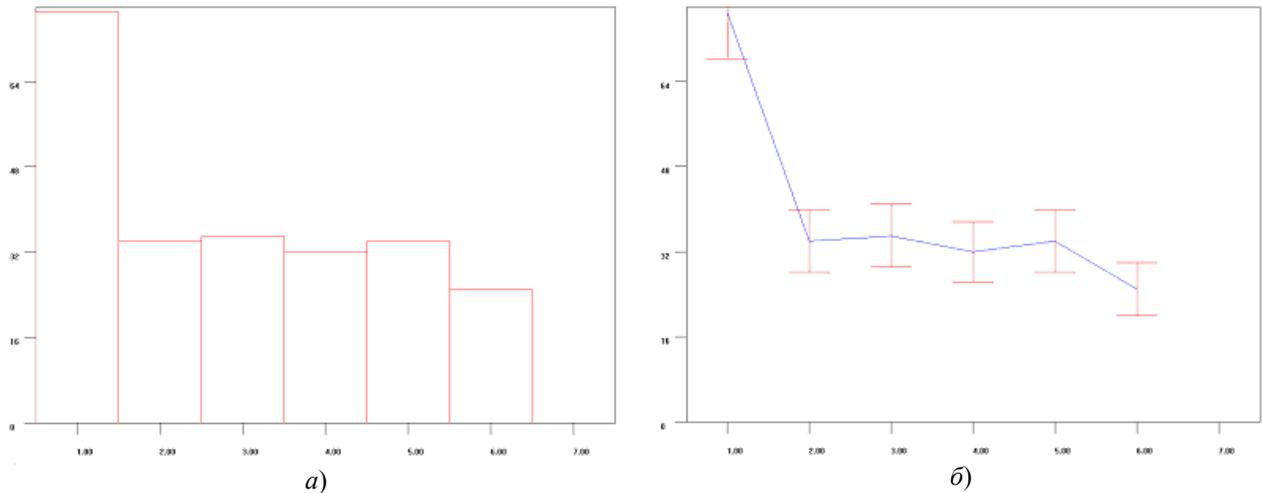

*а)* *б)*
Fig. 7. Distribution of intervals illustrating the effect of "near zone".

2) The synchronous, by local and absolute time, appearance of similar histograms at different geographical points is the second piece of evidence for the cosmo-physical conditionality for the histogram patterns in the time series of the GCP system.

The second, after the effect of "near zone", piece of evidence of the histogram pattern being determined by cosmo-physical factors is the dependence of this pattern on the rotation of the Earth around its axis.

We showed a high probability of similar histogram patterns to be realized synchronously by local time upon measurements of α-activity of $^{239}$Pu [3–13], noise in a gravitational antenna [3], dark current fluctuations in photomultipliers [7], rates of chemical reactions [13]. Simultaneously with measurements in our laboratory in Pushchino, there were measurements carried out aboard ships during expeditions in the Indian Ocean, in the Arctic and Antarctic; and in the laboratories of Russia (St. Petersburg, Moscow), Georgia (Tbilisi), Germany (Neuss, Lindau), Greece (Athens), Spain (Valencia), USA (Columbus). The local-time synchronism did not depend on the geographical latitude and the nature of the process studied and was observed with a high accuracy: with a resolution of 1 min at any distance between laboratories (up to 14000 km) [13].

It turns out that the time series obtained from generators of the GCP-system are not an exclusion. Similar histograms are observed synchronously by local time upon the GCP-measurements at different geographical points.

In a number of cases, we observed also an absolute-time synchronism upon study of physical processes, when similar histograms were highly probable to be realized at the same time.

However, such a global absolute synchronism appears differently in different experiments, and we failed to reveal a clear regularity in this phenomenon. Probably, the situation with GCP measurements is analogous. This is illustrated in Fig. 8-15.

Fig. 8 shows the dependence of the probability of similar histogram patterns to appear upon comparison of the results of measurements of June 7-8, 2000, which were taken from generators No. 28 (Roger Nelson, Princeton, NJ, USA; 40°350′ latitude north and 74°659′ longitude west) и No. 37 (John Walker, Switzerland; 47°079′ latitude north and 7°062′ longitude east). The local time difference is 327 min. As can be seen in Fig. 8, it is the time interval, corresponding to which is an extremum, namely the maximal probability of appearance of similar histograms constructed for the time series from those two generators. The right part of the same figure demonstrates the absence of any significant synchronism by absolute time.

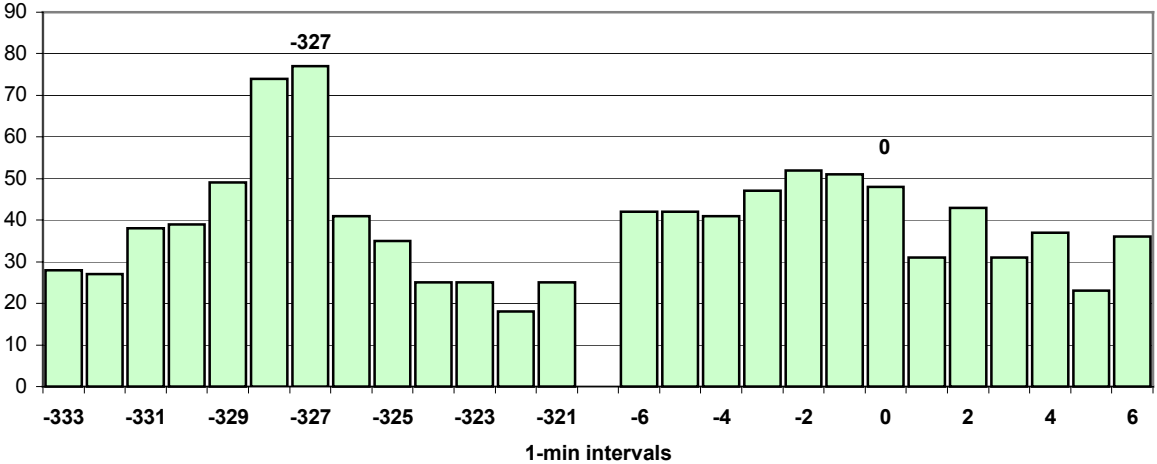

Fig. 8. The synchronism by local and absolute time. The dependence of the probability of similar histogram patterns to appear on the duration of the time interval between them. Compared are the results of measurements of June 7-8 2000, taken from generators No. 28 (Roger Nelson, Princeton, NJ, USA; 40°350′ latitude north and 74°659′ longitude west) and No. 37 (John Walker, Switzerland; 47°079′ latitude north and 7°062′ longitude east). At the left are intervals in the range of differences by local time. At the right are intervals in the range of differences by absolute time. Each histogram is constructed on the basis of a segment of the initial series equal to 1 min. The calculated local-time difference is 327 min.

Fig. 9 shows an analogous dependence obtained upon comparison of the results of measurements of April 8, 2005, taken from generators No. 37 (John Walker, Switzerland; 47°079′ latitude north and 7°062 longitude east) and no. 102 (Peter Mulacz, Wien, Austria; 48°217′ latitude north and 16°367′ longitude east). In this case, one can see a high probability of similar histograms to appear synchronously by local time, but there is also a notable synchronism by absolute time.

Fig. 10-12 give an additional illustration of the high probability of appearance of similar histograms constructed for the time series, which were produced by GCP generators at different geographical points at the same local time.

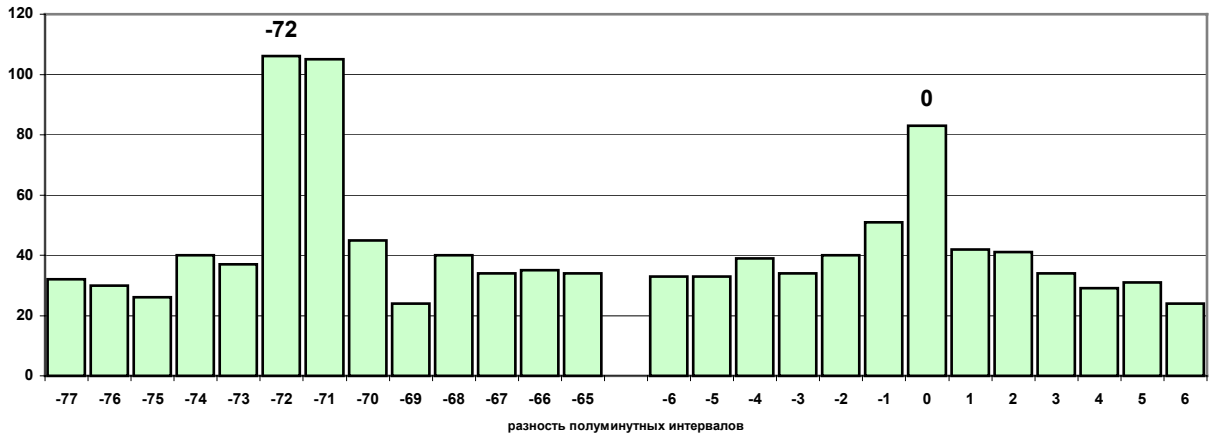

Fig. 9. The synchronism by local and absolute time. The dependence of the probability of similar histogram patterns to appear on the duration of the time interval between them. Compared are the results of measurements of April 8 2005, taken from generators No. 37 (John Walker, Switzerland; 47°079′ latitude north and 7°062′ longitude east) and No. 102 (Peter Mulacz, Wien, Austria; 48°217′ latitude north and 16°367′ longitude east). At the left are intervals in the range of differences by local time. At the right are intervals in the range of differences by absolute time. The duration of intervals is 0.5 min. The calculated local-time difference is 36 min (72 intervals).

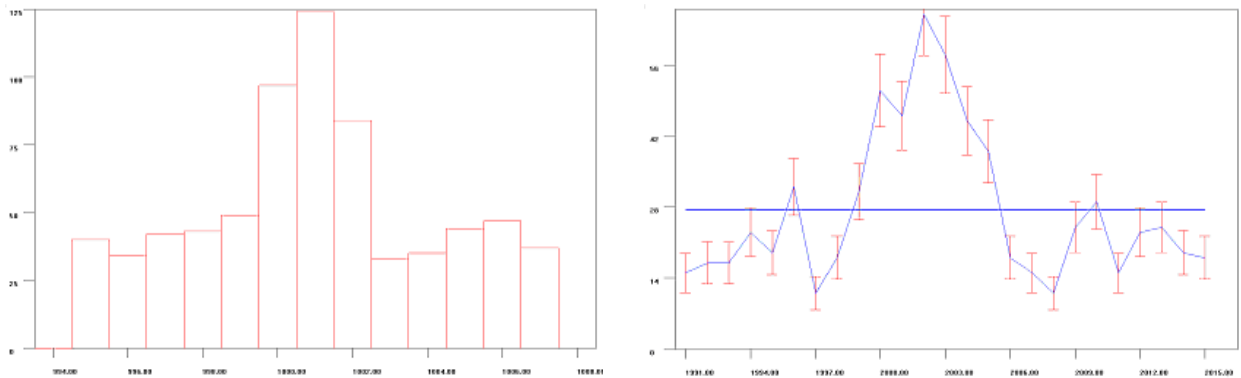

Fig. 10. The synchronism by local and absolute time. The dependence of the probability of similar histogram patterns to appear on the duration of the time interval between them. Compared are the results of measurements of April 8 2005, taken from generators No. 28 (Roger Nelson, Princeton, NJ, USA; 40°350′ latitude north and 74°659′ longitude west) and No. 100 (Robin Taylor, Simon Greaves, Suva, Fiji; 17°750′ latitude south and 177°45′ longitude east). At the left intervals are extended to 1 min. At the right are 0.5-min intervals. The root-mean-square error bars are given. The calculated local-time difference is 1006 min (2012 intervals).

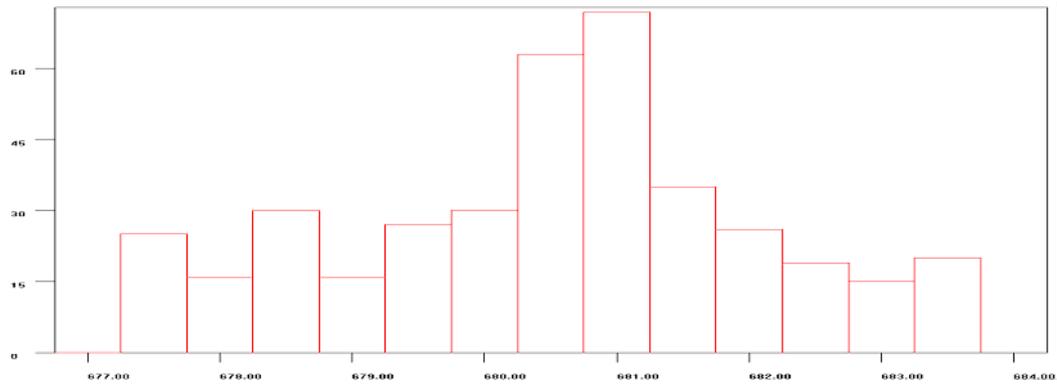

Fig. 11. The synchronism by local and absolute time. The dependence of the probability of similar histogram patterns to appear on the duration of the time interval between them. Compared are the results of measurements of April 8 2005, taken from generators No. 37 (John Walker, Switzerland; 47°079′ latitude north and 7°062′ longitude east) and No. 100 (Robin Taylor, Simon Greaves, Suva, Fiji; 17°750′ latitude south and 177°45′ longitude east). The calculated local-time difference is 681.8 min.

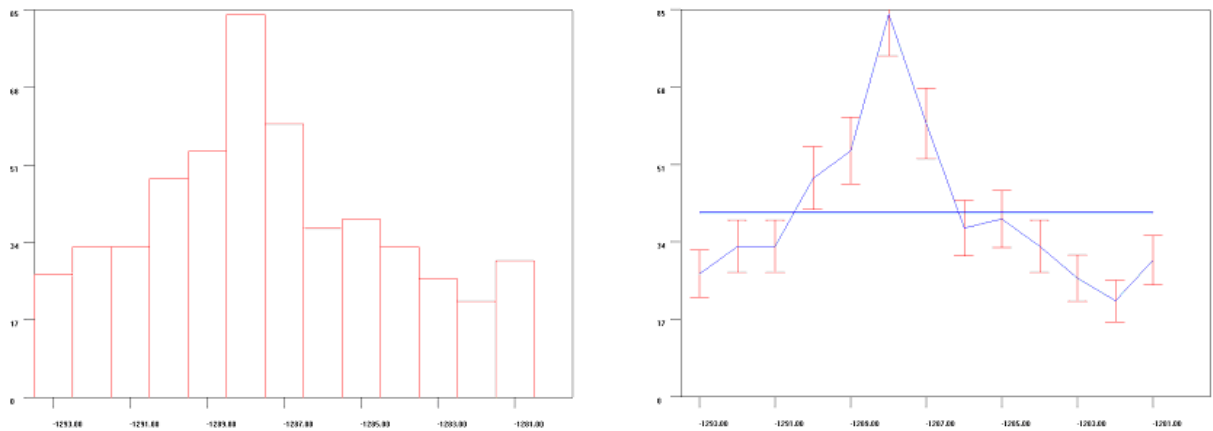

Fig. 12. The synchronism by local and absolute time. The dependence of the probability of similar histogram patterns to appear on the duration of the time interval between them. Compared are the results of measurements of April 8 2005, taken from generators No. 102 (Peter Mulacz, Wien, Austria; 48°217′ latitude north and 16°367′ longitude east) and No. 100 (Robin Taylor, Simon Greaves, Suva, Fiji; 17°750′ latitude south and 177°45′ longitude east). The calculated local-time difference is 644 min. The duration of intervals is 0.5 min.

The results presented indicate quite clearly that the regularities we found earlier in various physical processes are identical to those revealed in the time series produced by the GCP generators. However, it was psychologically important to make sure of this upon the direct comparison of histograms constructed from our typical results of measurements of $^{239}$Pu α-activity and histograms constructed on the basis of data from the GCP generators. The results of this comparison are shown in Fig. 13-14.

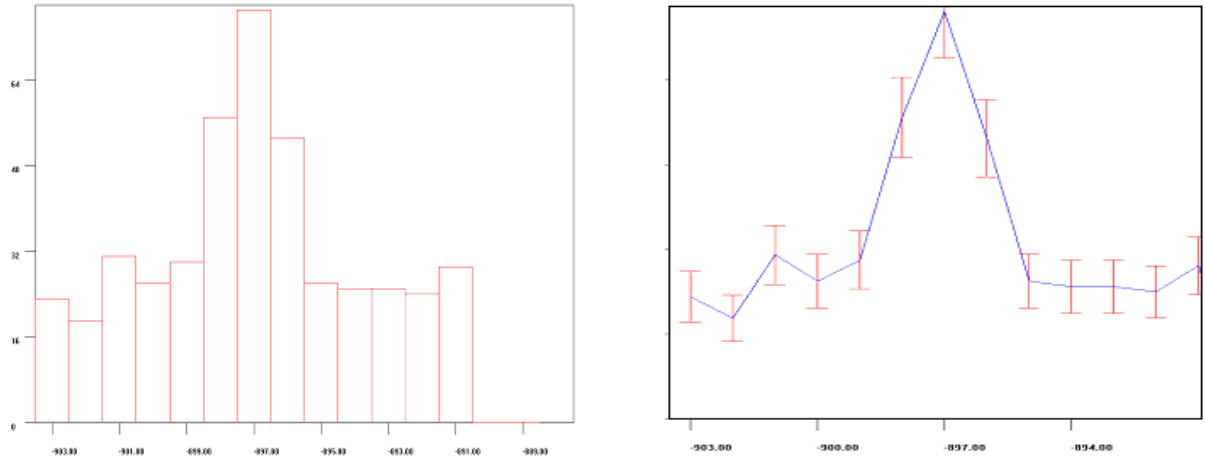

Fig. 13. The dependence of the probability of similar histogram patterns to appear on the duration of the time interval between them. Compared are the results of measurements of April 8 2005 taken from generator No. 102 (Peter Mulacz, Wien, Austria; 48°217′ latitude north and 16°367′ longitude east) and the results of measurements of α-activity of $^{239}$Pu in Pushchino (Simon Shnoll, Pushchino, Russia; 54°7′ latitude north and 37°6′ longitude east). At the right, the root-mean-square error bars are given. The duration of intervals is 0.5 min. The calculated local-time difference is 449 min.

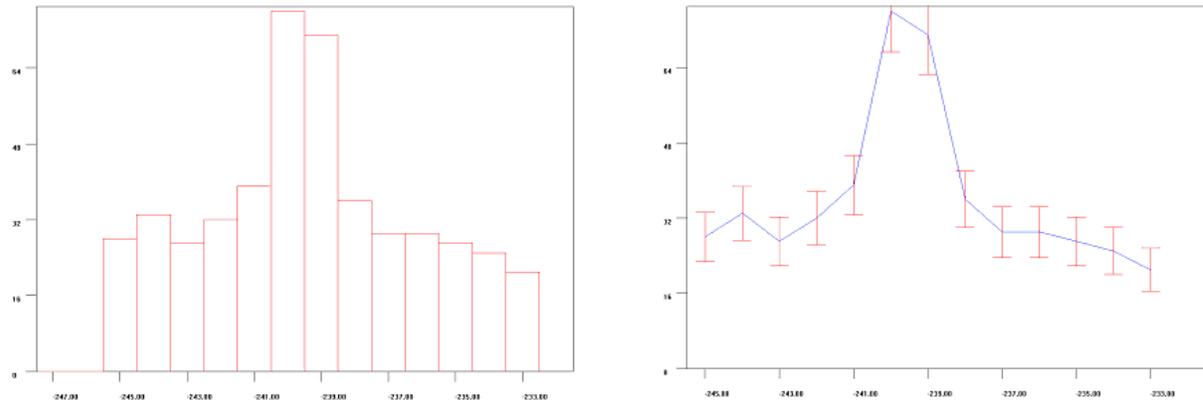

Fig. 14. The dependence of the probability of similar histogram patterns to appear on the duration of the time interval between them. Compared are the results of measurements of April 8 2005 taken from generator No. 37 (John Walker, Switzerland; 47°079′ latitude north and 7°062′ longitude east) and the results of measurements of α-activity of $^{239}$Pu in Pushchino (Simon Shnoll, Pushchino, Russia; 54°7′ latitude north and 37°6′ longitude east). At the right, the root-mean-square error bars are given. The duration of intervals is 0.5 min. The calculated local-time difference is 122 min.

3) The "sidereal" and "solar" daily periods in the change of the probability of recurring appearance of similar histogram patterns is the third piece of evidence for the cosmo-physical conditionality of histogram patterns in the time series of the GCP-system

As we have shown upon studying various physical processes, there are two clearly distinguishable daily periods in the change of the probability of histogram patterns to be similar: one of them is equal to the solar day (1440 min) and another is equal to the sidereal day (1436 min) [11]. As can be seen in Fig. 15, exactly the same periods are revealed in the sequences of histograms constructed for the time series from the GCP generators.

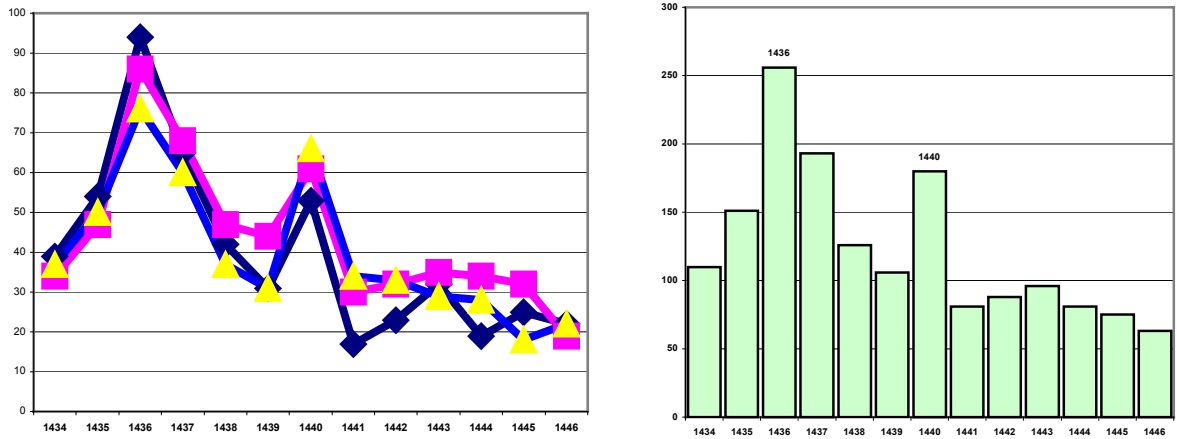

Fig. 15. The distribution of the number of pairs of similar 1-min histograms depending on the duration of the time interval between them (min). The probability of recurring appearance of similar histogram patterns varies with two different daily periods; they are equal to the "solar day" (1440 min) and the "sidereal day" (1436 min). Generators No. 28 and No. 37; at the right is a total distribution, at the left are results of three separate comparisons. Along the abscissa axis is the time interval between similar histograms (min). Along the axis of ordinates is the number of similar pairs corresponding to the given interval duration.

4) The synchronous realization of a characteristic histogram pattern at the moments of culminations of solar eclipses is the forth piece of evidence for the cosmo-physical conditionality of histogram patterns in the time series of the GCP-system

The data presented above were obtained by a traditional for our studies technique, i.e. by the pairwise comparison of histograms, calculation of the time interval between similar histograms, and construction of the distributions of the number of similar histogram pairs by the duration of the interval between histograms. This is a hard work. To construct each of the graphs given in the figures, we had to evaluate about 7000 of histogram pairs.

However, several years ago we found a cosmo-physical phenomenon, whose study does not require searching through dozens of histogram combinations. It turns out that at the moments of the new moon, a characteristic histogram pattern is realized with high probability practically simultaneously all over the Earth, at various geographical points from the Arctic to Antarctic, in the West and East hemispheres [9]. Recently, an analogous phenomenon has been discovered for the moments of solar eclipse culminations [14]. Let us leave the discussion of the physics of these amazing phenomena for a special analysis and note a change in the method, which is of principle importance. To reveal histogram patterns, specific for the new moon or solar eclipses, one need not to search for similar pairs through all the array of histograms. What we can do is to check at once is there a characteristic histogram pattern at a certain calculated moment. The application of this method to the series obtained from the GCP generators (with the purpose to analyse histograms, specific for culminations of solar eclipses) has confirmed the conclusion on the cosmo-physical conditionality of the corresponding histograms. The results are given in Fig. 16-24.

Fig. 16 demonstrates a fragment of the computer archive, a series of 1-min histograms constructed from the results of measurements of $^{239}$Pu α-activity in our Pushchino laboratory on October 3, 2005. Histogram No. 809 (colored red) corresponds to the eclipse culmination.

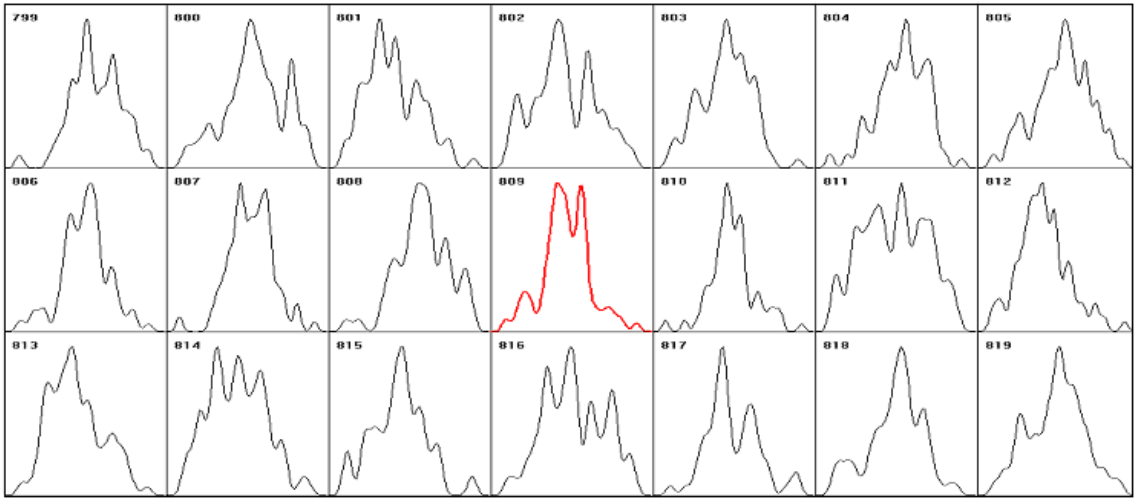

Fig. 16. A fragment of the computer archive, a series of 1-min histograms constructed from the results of measurements of $^{239}$Pu α-activity in our Pushchino laboratory during the solar eclipse on October 3, 2005 (Victor Panchelyuga, Pushchino, Russia; 54°7′ latitude north and 37°6′ longitude east). The histogram No. 809, having a characteristic pattern, appears 1 minute later of the calculated time (No. 808).

Fig. 17 shows an analogous fragment of the computer archive, a series of 0.5-min histograms constructed from the data of GCP generator No. 28 (Roger Nelson, Princeton, NJ, USA; 40°350′ latitude north and 74°659′ longitude west) collected during the solar eclipse of April 8, 2005. The calculated moment of the eclipse culmination and the appearance of the characteristic pattern coincide with the accuracy of 0.5 min (histogram No. 2486). Fig. 18 demonstrates a series of 0.5-min histograms constructed from the data of GCP generators No. 37 (John Walker, Switzerland; 47°079′ latitude north and 7°062′ longitude east) on April 8, 2005. No. 2486 corresponds to the calculated moment of eclipse culmination; the histogram of the characteristic pattern (No. 2487) differs from the calculated one by 0.5 min. Fig. 19 shows a series of 0.5-min histograms constructed from the data of GCP generator No. 100 (Robin Taylor, Simon Greaves, Suva, Fiji; 17 750′ latitude south and 177°45′ longitude east) of October 3, 2005. No. 1256 corresponds to the calculated moment of eclipse culmination; the histogram of the characteristic pattern appears exactly at this moment.

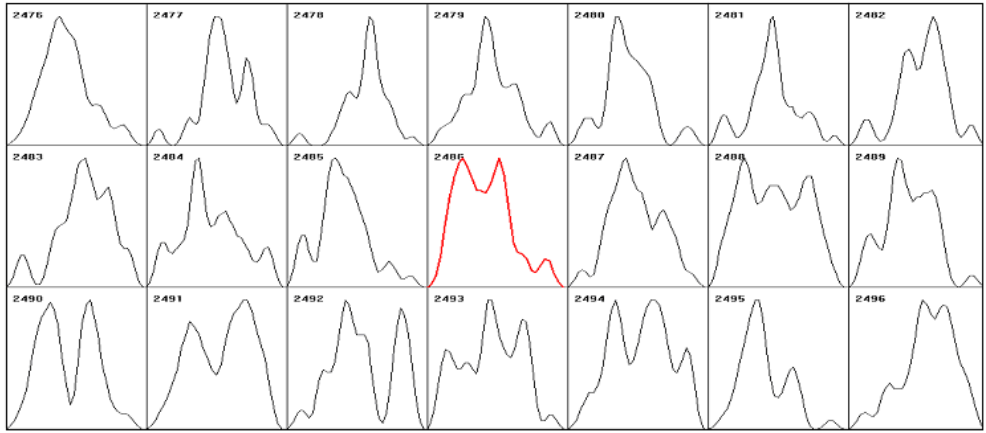

Fig. 17. A fragment of the computer archive, a series of 0.5-min histograms constructed from the data of GCP generator No. 28 (Roger Nelson, Princeton, NJ, USA; 40°350′ latitude north and 74°659′ longitude west) during the solar eclipse of April 8, 2005. The calculated moment of the eclipse culmination and the appearance of a characteristic pattern coincide with an accuracy of 0.5 min (histogram No. 2486).

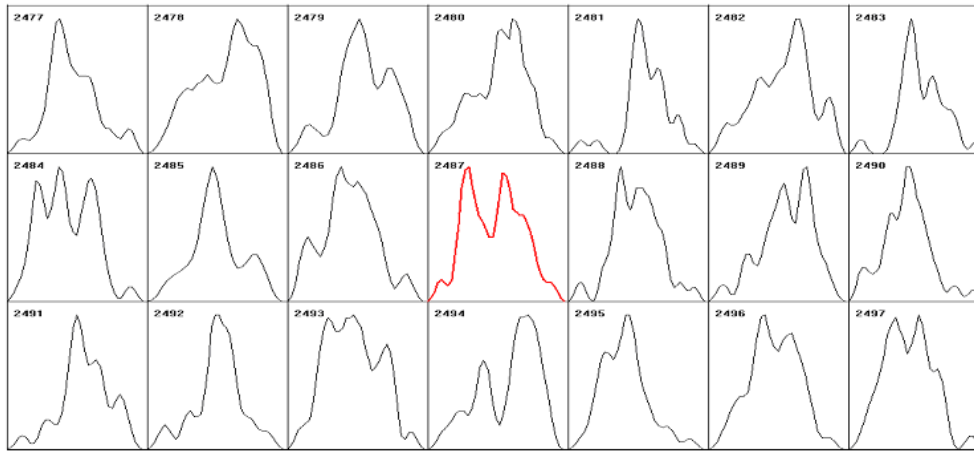

Fig. 18. A series of 0.5-min histograms constructed from the data of GCP generator No. 37 (John Walker, Switzerland; 47°079′ latitude north and 7°062′ longitude east) of April 8, 2005. No. 2486 corresponds to the calculated moment of eclipse culmination; histogram No. 2487, having a characteristic pattern, differs from that at the calculated moment by 0.5 min.

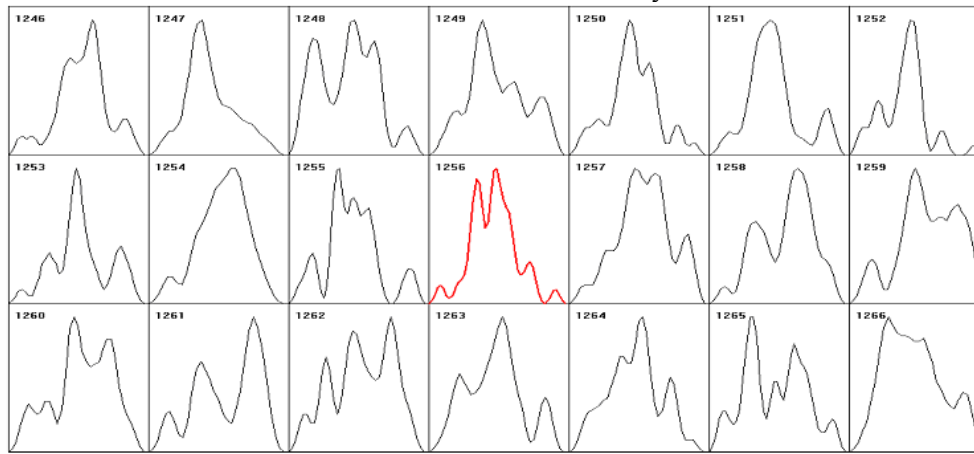

Fig. 19. A series of 0.5-min histograms constructed from the data of GCP generator No. 100 (Robin Taylor, Simon Greaves, Suva, Fiji; 17°750′ latitude south and 177°45′ longitude east) of October 3, 2005. No. 1256 corresponds to the calculated moment of eclipse culmination; the histogram of the characteristic pattern is realized exactly at this moment.

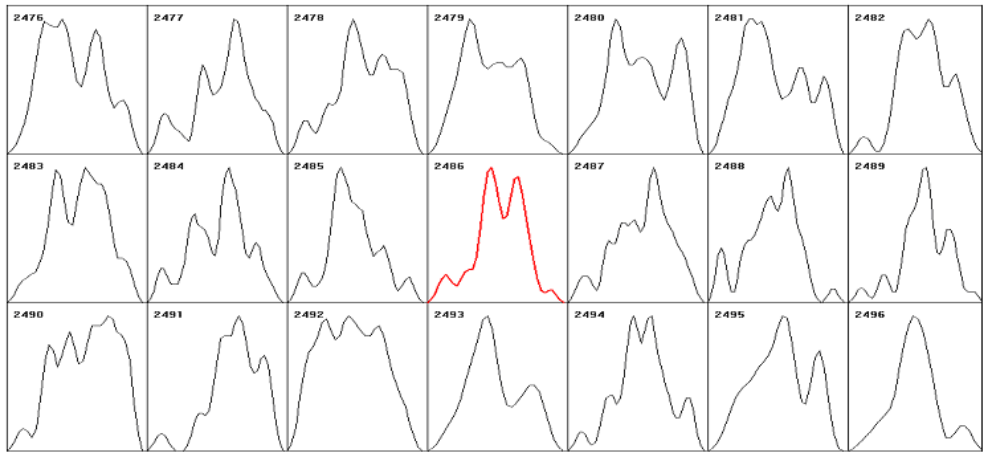

Fig. 20. A series of 0.5-min histograms constructed from the data of GCP generator No. 103 (Rick Berger, San Antonio, TX, USA; 29°493′ latitude north and 98°612′ longitude east) of April 8, 2005. No. 2486 corresponds to the calculated moment of eclipse culmination; the histogram of the characteristic pattern is realized exactly at this moment.

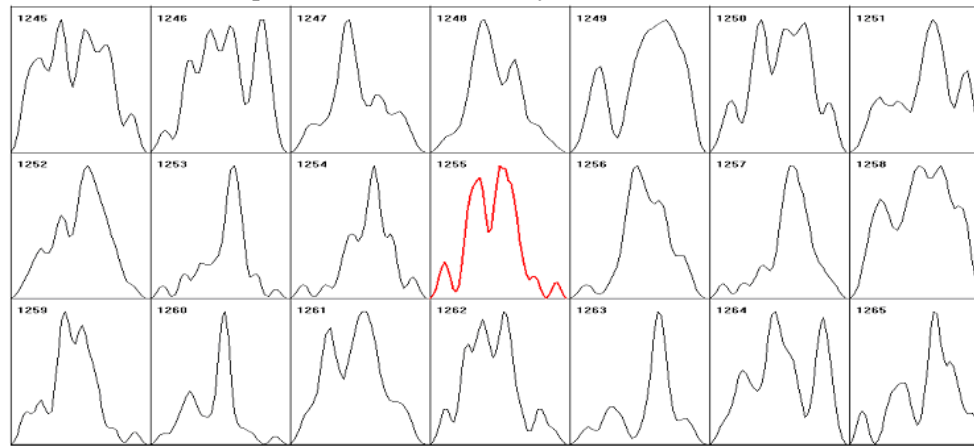

Fig. 21. A series of 0.5-min histograms constructed from the data of GCP generator No. 28 (Roger Nelson, Princeton, NJ, USA; 40°350′ latitude north and 74°659′ longitude west) of October 3, 2005. No. 1256 corresponds to the calculated moment of eclipse culmination; histogram No. 1255, having a characteristic pattern, is realized 0.5 min earlier of that moment.

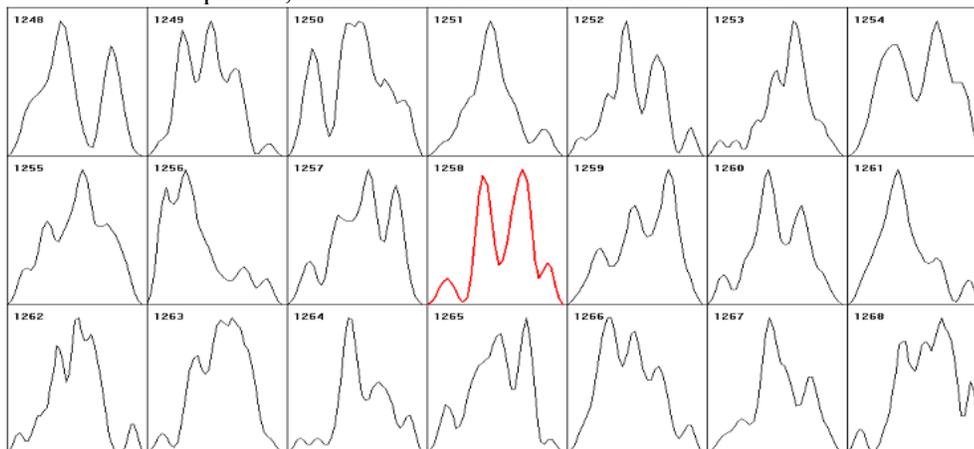

Fig. 22. A series of 0.5-min histograms constructed from the data of GCP generator No. 37 (John Walker, Switzerland; 47°079′ latitude north and 7°062′ longitude east) of October 3, 2005. No. 1256 corresponds to the calculated moment of eclipse culmination; histogram No. 1258, having a characteristic pattern, is realized 1 min later of that moment.

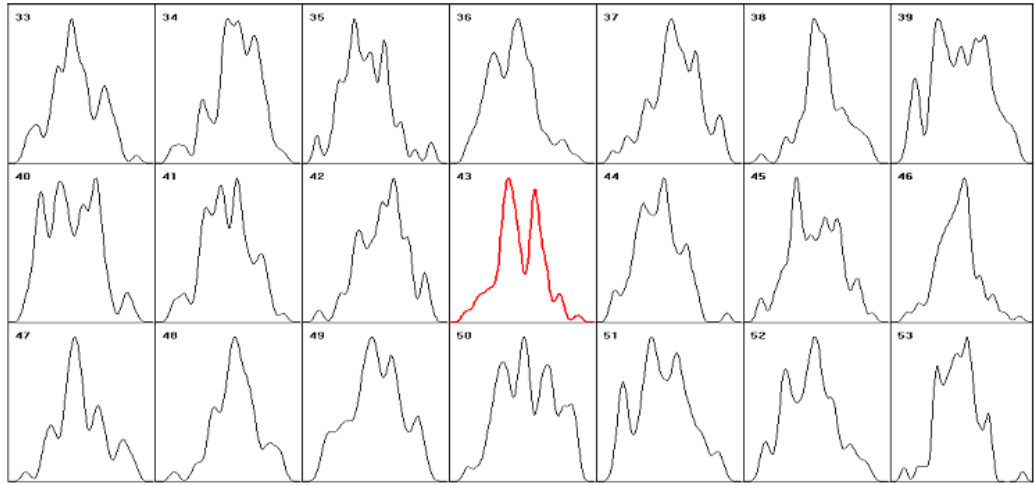

Fig. 23. A fragment of the computer archive, a series of 1-min histograms constructed from the results of measurements of $^{239}$Pu α-activity in our Pushchino laboratory (Simon Shnoll, Pushchino, Russia; 54°7′ latitude north and 37°6′ longitude east) during the solar eclipse on April 9 (8), 2005. The histogram of the characteristic pattern (No. 43, red-colored) is realized exactly at the calculated time.

The given series of figures shows that the histograms of the specific for the culmination of the solar eclipse pattern are realized exactly at the calculated time – both upon the measurements of radioactivity and upon the data generation in the GCP-network. Therefore, this parameter also reflects the cosmo-physical nature of the factors that determine the fine structure (pattern) of histograms.

## 4. NUMERICAL SIMULATION OF THE ALGORITHM FOR PROCESSING OF DATA OBTAINED FROM THE PHYSICAL RANDOM NUMBER GENERATORS USED IN THE GCP-NETWORK.

As noted in the Introduction, one of the questions raised in connection with the use of the GCP-network data is the following: how much does the application of the XOR-mask "corrupt" data obtained from a physical random number generator? In other words, wouldn't a physical random number generator be transformed into a logical one, such as a computer random number generator, whose output data will depend not on a physical process but on the algorithm applied? The data presented above suggest a negative answer to this question. Nevertheless, we consider necessary to examine the results of a numerical simulation of the effect of XOR-mask on the histogram pattern.

The numerical model, which is designed to examine the possibility of using the GCP-network data for study of the effect of macroscopic fluctuations, simulate the process of retrieval and pre-processing of experimental data by the physical generators used in the GCP-network. The working algorithm of this numerical model is given below.

Fig. 24*a* represents an initial signal, analogous to the signal from a physical random number generator after a low-pass filter. This signal is transformed into a meander according to the following algorithm:

(1)
$$S_{out} = \begin{cases} 1 : S_{in} > 0, \\ 0 : S_{in} \leq 0. \end{cases}$$

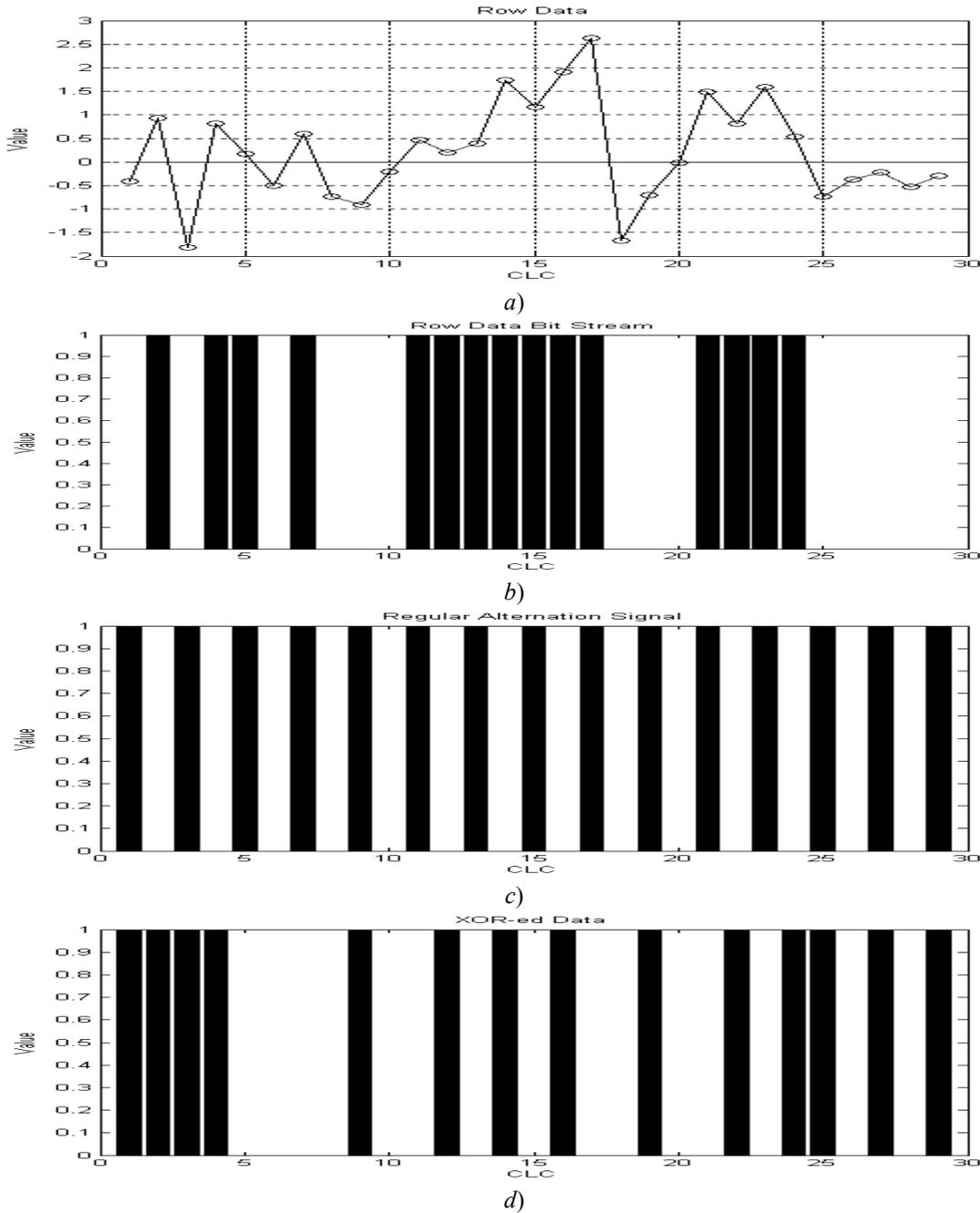

Fig. 24. The algorithm for experimental data processing used to simulate the work of generators of the GCP-network.

The results of transformation of the input sequence (Fig. 24*a*) into the sequence of bits according to (1) is given in Fig. 24*b*. The algorithm (1) allows one to transform an analog signal into a digital binary sequence. This, in its turn, allows one to apply to the output sequence a mask of "exclusive OR". To do this, one should prepare a sequence of periodically alternating zeros and unities:

(2) $$S_{01} = \{0,1,0,1,0,1,...\},$$

which is given in Fig. 24*c*. The sequences $S_{01}$ and $S_{out}$ have the same length. The operation of "exclusive OR" is realized according to the following algorithm:

(3) $$S_{XOR} = \begin{cases} 1 : S_{out} \neq S_{01} \\ 0 : S_{out} = S_{01} \end{cases}.$$

The result of applying the algorithm (3) to the sequences shown in Fig. 24*b* and 24*c* is given in Fig. 24*d*.

As noted above, a single point of the output signal of the physical random number generator is a result of summation of the number of unities over 200 beats of the $S_{XOR}$ signal. Therefore, to get an output sequence of 1000-point length, one needs sequences $S_{out}$ and $S_{01}$ of 200000-point length. Fig. 25 demonstrates two sequences, $S_{out}^{200}$ and $S_{XOR}^{200}$, every point of which is obtained by summation of 200 points of $S_{out}$ and $S_{XOR}$. It is necessary to note that every point of the sequence $S_{out}^{200}$ is also obtained by summation of 200 points of the sequence $S_{out}$.

The signals $S_{out}^{200}$ and $S_{XOR}^{200}$ given in Fig. 25 were further examined by the technique of histogram analysis briefly described above. The examination revealed that the application of the XOR-mask preserves the main extrema that are characteristic for histograms of the sequence $S_{out}^{200}$ and only slightly distorts the histogram shape. This fact is in agreement with the data presented above, which demonstrate the main laws of the phenomenon of macroscopic fluctuations. At the same time, the distortions caused by the mask can make impossible the distinction of complex histogram patterns. Thus, for example, analysing the time series from the GCP generators, we failed to reveal the histogram patterns characteristic for the new moon [9] (they have a more complex structure than the "solar eclipse" patterns presented in this work). Nevertheless, as the extrema coincide, we can hope that the use of methods of statistical manipulation, which is possible owing to a quite large length of the GCP series, will allow us to distinguish rather complex patterns.

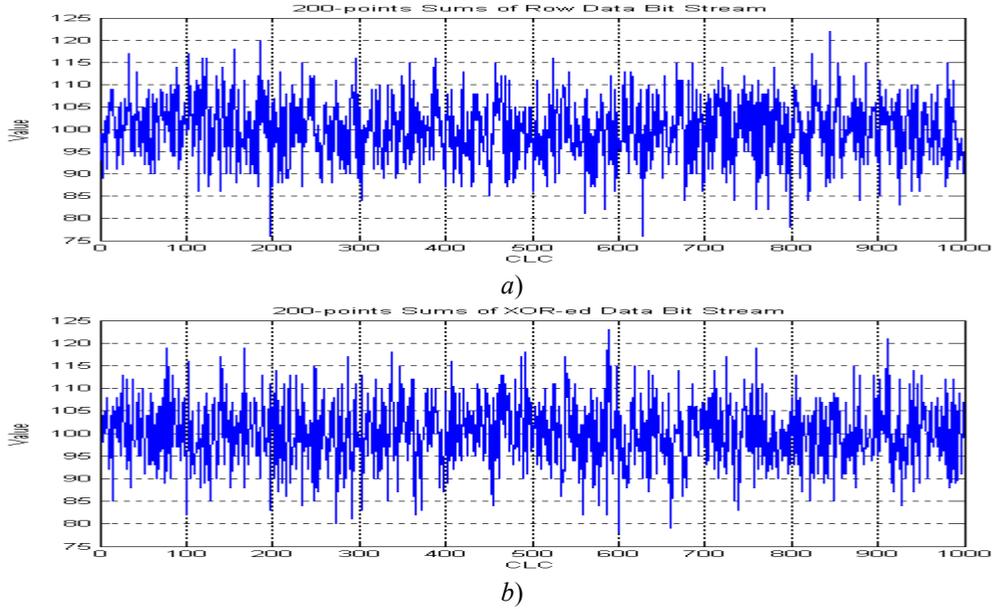

Fig. 25. The signals $S_{out}^{200}$ (*a*) and $S_{XOR}^{200}$ (*b*) at the output of the model implementing the algorithm of the GCP data processing.

## 4. Discussion

As follows from the data presented, the time series produced by the generators of the GCP-system reveal all the main signs of "macroscopic fluctuations", i.e. the regular changes in the fine structure of the spectrum of fluctuations or, in other words, in the pattern of the corresponding histograms. Therefore, the pattern of these histograms is determined by cosmo-physical factors, resulting from the spatial anisotropy and temporal heterogeneity of our universe [5, 6 13]. The application of the XOR-mask does not remove the dependence on physical factors in the time series obtained from the generators of the GCP-network.

We do not consider our conclusions as contradicting the results obtained within the scope of the GCP-project, since they are based on such methods of analysis of time series, which differ, in principle, from those used in this project. At the same time, we hope that our results may be useful for the GCP-community.

It should be noted that the long-term measurements in the GCP-system are exceptionally valuable. They may contain a great deal of information on cosmo-physical factors and processes, which can be revealed by the method of histogram analysis.

In this paper, we almost do not concern questions on the nature of the effects observed. The hypotheses and attempts of theoretical interpretation of the phenomena observed, which have been proposed for many years, are summarized in the review [13].